\documentclass[12pt,english]{article}

\usepackage{algorithm}
\usepackage{booktabs}
\usepackage{graphicx}
\usepackage{amsmath}
\usepackage{amssymb}
\usepackage{latexsym}
\usepackage{crop}
\usepackage{algorithmic,algorithm}
\usepackage{multirow}
\usepackage{bm}
\usepackage{bbm}
\usepackage{enumerate}
\usepackage{url}
\usepackage{array}
\usepackage{paralist}
\usepackage{diagbox}
\usepackage{wasysym}
\usepackage{booktabs}
\usepackage[dvipsnames]{xcolor}
\usepackage[colorlinks = true, pdfstartview = FitV, linkcolor = blue, citecolor = blue, urlcolor = blue]{hyperref}

\usepackage{listings}

\usepackage{rotating}

\usepackage[capitalise]{cleveref}
\crefname{equation}{}{}
\crefname{figure}{Figure}{Figures}
\creflabelformat{equation}{\textup{(#2#1#3)}}
\crefname{assumption}{Assumption}{Assumptions}
\crefname{condition}{Condition}{Conditions}
\usepackage{xspace}

\usepackage{fullpage}
\usepackage{multirow}

\usepackage[sort,nocompress]{cite}
\usepackage{arydshln}
\usepackage{enumitem}
\setlist[enumerate,1]{leftmargin=*,wide=0em, noitemsep,nolistsep, label = {\bfseries \arabic*.}}
\setlist[itemize,1]{leftmargin=*,wide=0em, noitemsep,nolistsep}
\usepackage{pifont}% http://ctan.org/pkg/pifont
\makeatletter
\newcommand*{\transpose}{%
	{\mathpalette\@transpose{}}%
}
\newcommand*{\@transpose}[2]{%
	\raisebox{\depth}{$\m@th#1\intercal$}%
}
\makeatother
\newcommand*\xbar[1]{%
	\hbox{%
		\vbox{%
			\hrule height 0.5pt % The actual bar
			\kern0.5ex%         % Distance between bar and symbol
			\hbox{%
				\kern-0.1em%      % Shortening on the left side
				\ensuremath{#1}%
				\kern-0.1em%      % Shortening on the right side
			}%
		}%
	}%
} 

\definecolor{forestgreen}{rgb}{0.13, 0.55, 0.13}

\definecolor{amber}{rgb}{1.0, 0.75, 0.0}

\definecolor{bananayellow}{rgb}{.8, 0.6, 0}

\newcounter{comment}

\usepackage{amsthm}
\usepackage[framemethod=TikZ]{mdframed}

\mdfdefinestyle{theoremstyle}{%
	linewidth = 1pt,%
	roundcorner = 10pt,%
	leftmargin = 0,%
	rightmargin = 0,%
	backgroundcolor = cyan!3,%
	outerlinecolor = magenta!70!black,%
	splittopskip = \topskip,%
	ntheorem = true,%
}
\mdtheorem[style=theoremstyle]{claim}{Claim}

\newmdtheoremenv[%
linewidth = 1pt,%
roundcorner = 10pt,%
leftmargin = 0,%
rightmargin = 0,%
backgroundcolor = green!3,%
outerlinecolor = blue!70!black,%
splittopskip = \topskip,%
ntheorem = true,%
]{theorem}{Theorem}

\newmdtheoremenv[%
linewidth = 1pt,%
roundcorner = 10pt,%
leftmargin = 0,%
rightmargin = 0,%
backgroundcolor = green!3,%
outerlinecolor = blue!70!black,%
splittopskip = \topskip,%
ntheorem = true,%
]{corollary}{Corollary}

\newmdtheoremenv[%
linewidth = 1pt,%
roundcorner = 10pt,%
leftmargin = 0,%
rightmargin = 0,%
backgroundcolor = green!3,%
outerlinecolor = blue!70!black,%
splittopskip = \topskip,%
ntheorem = true,%
]{lemma}{Lemma}

\newmdtheoremenv[%
linewidth = 1pt,%
roundcorner = 10pt,%
leftmargin = 0,%
rightmargin = 0,%
backgroundcolor = yellow!3,%
outerlinecolor = blue!70!black,%
splittopskip = \topskip,%
ntheorem = true,%
]{definition}{Definition}

\newmdtheoremenv[%
linewidth = 1pt,%
roundcorner = 10pt,%
leftmargin = 0,%
rightmargin = 0,%
backgroundcolor = green!3,%
outerlinecolor = blue!70!black,%
splittopskip = \topskip,%
ntheorem = true,%
]{proposition}{Proposition}

\newmdtheoremenv[%
linewidth = 1pt,%
roundcorner = 10pt,%
leftmargin = 0,%
rightmargin = 0,%
backgroundcolor = green!3,%
outerlinecolor = blue!70!black,%
splittopskip = \topskip,%
ntheorem = true,%
]{condition}{Condition}

\newmdtheoremenv[%
linewidth = 1pt,%
roundcorner = 10pt,%
leftmargin = 0,%
rightmargin = 0,%
backgroundcolor = green!3,%
outerlinecolor = blue!70!black,%
splittopskip = \topskip,%
ntheorem = true,%
]{assumption}{Assumption}

\theoremstyle{definition}
\newmdtheoremenv[%
linewidth = 1pt,%
roundcorner = 10pt,%
leftmargin = 0,%
rightmargin = 0,%
backgroundcolor = blue!3,%
outerlinecolor = blue!70!black,%
splittopskip = \topskip,%
ntheorem = true,%
]{example}{Example}

\theoremstyle{definition}
\newmdtheoremenv[%
linewidth = 1pt,%
roundcorner = 10pt,%
leftmargin = 0,%
rightmargin = 0,%
backgroundcolor = red!3,%
outerlinecolor = blue!70!black,%
splittopskip = \topskip,%
ntheorem = true,%
]{remark}{Remark}

\usepackage{tikz}
\usepackage{xparse}% So that we can have two optional parameters

\NewDocumentCommand\DownArrow{O{2.0ex} O{black}}{%
	\mathrel{\tikz[baseline] \draw [<-, line width=0.5pt, #2] (0,0) -- ++(0,#1);}
}

\usepackage{listings} % to inser code

\definecolor{mygreen}{rgb}{0,0.6,0}
\definecolor{mygray}{rgb}{0.5,0.5,0.5}
\definecolor{mymauve}{rgb}{0.58,0,0.82}

\usepackage{dsfont}

\usepackage[latin1]{inputenc}
\usepackage{amsmath}
\usepackage{amsfonts}
\usepackage{amssymb}
\usepackage{makeidx}
\usepackage{graphicx}
\usepackage{caption}
\usepackage{subcaption}
\usepackage{framed}
\usepackage{booktabs,array}
\usepackage{xcolor}

\allowdisplaybreaks

\begin{document}
\title{The Importance of Environmental Factors in Forecasting Australian Power Demand}
\author{
	Ali Eshragh\thanks{School of Information and Physical Sciences, University of Newcastle, NSW, Australia, and International Computer Science Institute, Berkeley, CA, USA. Email:  \tt{ali.eshragh@newcastle.edu.au}}
	\and
	Benjamin Ganim\thanks{School of Information and Physical Sciences, University of Newcastle, NSW, Australia. Email:  \tt{Benjamin.Ganim@uon.edu.au}} 
	\and 
	Terry Perkins\thanks{School of Information and Physical Sciences, University of Newcastle, NSW, Australia. Email:  \tt{Terry.Perkins@uon.edu.au}} 
	\and 
	Kasun Bandara\thanks{School of Computing and Information Systems, Melbourne Centre for Data Science, University of Melbourne, VIC, Australia. Email: \tt{Kasun.Bandara@unimelb.edu.au}}
}
\date{}
\maketitle

\begin{abstract}
	We develop a time series model to forecast weekly peak power demand for three main states of Australia for a yearly time-scale, and show the crucial role of environmental factors in improving the forecasts. More precisely, we construct a seasonal autoregressive integrated moving average (\texttt{SARIMA}) model and reinforce it by employing the exogenous environmental variables including, maximum temperature, minimum temperature, and solar exposure. The estimated hybrid \texttt{SARIMA}-regression model exhibits an excellent mean absolute percentage error (MAPE) of $3.41\%$. Moreover, our analysis demonstrates the importance of the environmental factors by showing a remarkable improvement of $46.3\%$ in MAPE for the hybrid model over the crude \texttt{SARIMA} model which merely includes the power demand variables. In order to illustrate the efficacy of our model, we compare our outcome with the state-of-the-art machine learning methods in forecasting. The results reveal that our model outperforms the latter approach.
\end{abstract}

%----------------------
\section{Introduction}
\label{SecInt}
%----------------------

Electrical energy is a vital resource to drive industries \cite{SA2010}. Thus, energy demand forecasting is essential to the economic and socioeconomic aspects of modern society. Accurate forecasts ensure that utilities can meet energy demand and avoid undesirable events in the network such as black-outs and load shedding. While underestimation is undesirable, overestimation leads to wasted resources. In spite of recent advances in storage technologies, demand forecasting models are still critical in power planning \cite{CHIK2012}. 

In general, there are four main time-scales (or, forecast horizons) for power demand modeling \cite{HERN2014}:

\begin{enumerate}[label = (\roman*)]
	\item Long-term load forecasting (LTLF) is used for expansion planning of the network;
	
	\item Medium-term load forecasting (MTLF) is used for operational planning;
	
	\item Short-term load forecasting (STLF) is used for day to day planning and dispatch cost minimization;
	
	\item Very short-term load forecasting (VSTLF) on the scale of seconds to minutes allows the network to respond to the flow of demand.
\end{enumerate}

Australia is a vast and environmentally diverse continent with climate zones ranging from equatorial to temperate. It is thus important to understand how the dynamics of power demand varies across different regions. 

In this paper, we develop a seasonal autoregressive integrated moving average (\texttt{SARIMA}) model to forecast peak weekly demand in the medium-term (i.e., MTLF). The demand data are from three main Australian states consisting of: New South Wales (NSW), Victoria (VIC), and South Australia (SA). To investigate the impact of the environmental factors on the power demand, we hybridize the \texttt{SARIMA} model with a linear regression model by employing the exogenous environmental variables including, maximum temperature, minimum temperature, and solar exposure. Our results reveal that the latter hybrid model improves the accuracy of forecasts by an average factor of $46.3\%$ over the three states. Furthermore, to demonstrate the efficacy of the hybrid model, its outputs are compared with the state-of-the-art machine learning methods in forecasting. The results reveal that the former hybrid model outperforms the latter methods. 

The structure of this paper is organized as follows: Section \ref{SecLtR} provides a review of the literature and establishes the motivation for using a \texttt{SARIMA}-regression model. Section \ref{SecData} discusses the data resources and aggregation, and visualizes the obtained time series. Section \ref{SecSARIMA} explains the details of the statistical procedure to fit a \texttt{SARIMA} model to the weekly peak power demand data. In Section \ref{SecWV}, we employ secondary environmental time series to construct a hybrid \texttt{SARIMA}-regression model. Section \ref{SecFore} discusses the quality of $52$-week forecasts and compare the outcome with the state-of-the-art machine learning methods in forecasting. Finally, Section \ref{SecCon} presents a final discussion of our findings, and provides conclusions and directions for future research.

%------------------------------------------
\section{Literature Review and Motivation}
\label{SecLtR}
%------------------------------------------

Energy demand is an amalgamation of millions of individual demand requirements from consumers, varying with time, weather, population growth, electricity price and many other economic factors (e.g., see \cite{ZHU2011} and \cite{KARE2006}). The time dependency of the demand along with its inherent seasonality to weather patterns across a yearly time-scale would suggest time series methods to study the dynamics of the demand. 

Box and Jenkins \cite{BOX2015} introduced their celebrated 
\texttt{SARIMA} model for analyzing those non-stationary time series displaying seasonal effects in their behavior. Each \texttt{SARIMA} model is a linearly transformed time series constructed by differencing the original time series at proper lags. A hybrid \texttt{SARIMA}-regression approach could be effective, if the time covariance of the series is well captured by the \texttt{SARIMA} component and the remaining mean value of trends is captured by the exogenous independent variables (e.g., see \cite{Abol2020,CHIK2012}).  Although it has been more than $40$ years since such model were developed, due to their simplicity and vast practicality, they continue to be widely used in theory and practice, particularly effectively in electricity demand forecasting. 

Crude \texttt{SARIMA} as well as hybrid \texttt{SARIMA}-regression models have formed the basis of many power forecasting models with a focus on STLF to MTLF time-scale (i.e., looking days to weeks ahead) in several countries, as Nigeria \cite{MA2009}, Iraq \cite{KARE2006}, Malaysia \cite{MO2010}, South Africa \cite{CHIK2012}, and Thailand \cite{KA2011}. Focusing on a metric of peak demand ensures that demand can be met when the electricity network is under maximum duress. Ghalehkhondabi et al. \cite{GH2010} studied the peak monthly demand in Northern India by using two different time series methods including ``\texttt{SARIMA}'' and ``exponential smoothing'' models. The authors showed that the \texttt{SARIMA} model outperformed the exponential smoothing model on their data. In Australia, Amaral et al. \cite{AMAR2008} developed a smooth transition periodic autoregressive model for the New South Wales power demand, and As'ad \cite{AS2012} predicted the peak demand for New South Wales at a daily resolution. For a more comprehensive overview of such techniques in power demand modeling and forecasting, see \cite{GH2010}. 

In time series forecasting, global forecasting methods (GFM) that simultaneously learns from a collection of time series, are becoming a strong alternative to the state-of-the-art univariate statistical forecasting method such as \texttt{SARIMA} \cite{Smyl2019-cb,Bandara2019-iv}. In GFMs, a unified model is built using a set of related time series that enables the model to exploit key structures, behaviors, and patterns common within a group of time series. In fact, more recently, \emph{deep learning} based GFMs have shown promising results in forecasting competitions and real-world applications (e.g., see \cite{Bandaramultiple2020-il,Smyl2019-cb,Bandara2019-iv,Flunkert2017-fa,BandaraPPM2020-il}).

While \emph{artificial neural networks} (ANN) are increasing in popularity, Kandananond \cite{KA2011} compared ANN, multiple linear regression (MLR) and \texttt{SARIMA} models for electricity demand forecasting in Thailand. Although, they did not find a statistically significant difference between the three methods,  MLR and \texttt{SARIMA} were simpler to compute, and the coefficients were more easily interpreted. 

In this paper, we develop a hybrid \texttt{SARIMA}-regression model to forecast the weekly peak power demand in Australia over an MTLF time-scale, that is one year horizon ($52$ weeks). The main contribution of this work is to demonstrate the crucial role of novel environmental variables in the dynamics of the demand. The quality of forecasts are compared with the state-of-the-art machine learning techniques. The results show that our model not only outperforms the others, but also can more easily be computed and interpreted. 

We conclude this section by noting that as electricity energy is still difficult to store, it is critical that the system can meet peak demand \cite{ZHU2011}. To the best of our knowledge, this work is the first attempt to investigate the impact of \emph{environmental factors} on predicting the aggregated \emph{weekly peak demand} in an MTLF time-scale study.

%----------------------------------------------------
\section{Data: Resources, Aggregation and Visualizing}
\label{SecData}
%---------------------------------------------------

The power demand data for three major states of Australia, consisting of New South Wales (NSW), Victoria (VIC), and South Australia (SA), are obtained from the Australian Energy Market Operator \cite{AEMO2018}. They are measured in megawatts (MW). The secondary environmental time series data are acquired from the Australian Bureau of Meteorology \cite{ABM2018}. We use the data from those weather stations in close proximity to the primary population center for each state. These major population centers are Sydney, Melbourne, and Adelaide for NSW, VIC, and SA, respectively. Table \ref{sites} lists the details of those weather stations. 
\begin{table}[h!]
 \centering
	\caption{Australian Bureau of Meteorology weather stations}
	\label{sites}       % Give a unique label
	\begin{tabular}{lll}
		\hline\noalign{\smallskip}
		State & Site & BoM Site Number   \\
		\noalign{\smallskip}\hline\noalign{\smallskip}
		NSW & Sydney Airport & 066037  \\
		VIC & Melbourne Airport & 086282 \\
		SA & Brisbane Weather Station & 040913  \\
		\noalign{\smallskip}\hline
	\end{tabular}
\end{table}

While the power demand data are given at $15$-minute intervals, the environmental data are recorded weekly. So the former are aggregated by finding the peak demand for each day and then aggregating on a weekly basis. This aggregated value will be referred to as the \emph{weekly peak demand} (WPD). The weekly data from the first week of January $2011$ to the last week of December $2016$ (i.e., six years) are used as the training data for modeling and estimating the parameters. Following the MTLF time-scale, the data from the first week of January $2017$ to the last week of December $2017$ (i.e., $52$ weeks) are used as the test data to check the accuracy of forecasts generated by the model. 

The three secondary environmental time series used in this work are ``maximum temperature'', ``minimum temperature'', and ``solar exposure'', denoted by $\mathtt{Min}_t$, $\mathtt{Max}_t$ and $\mathtt{Sol}_t$, respectively. \emph{Solar exposure} is defined as the amount of solar energy falling on a flat one meter square surface, parallel to the ground and exposed to  direct sunlight.

\begin{figure}[h]
	\begin{center}
		\includegraphics[scale=0.50]{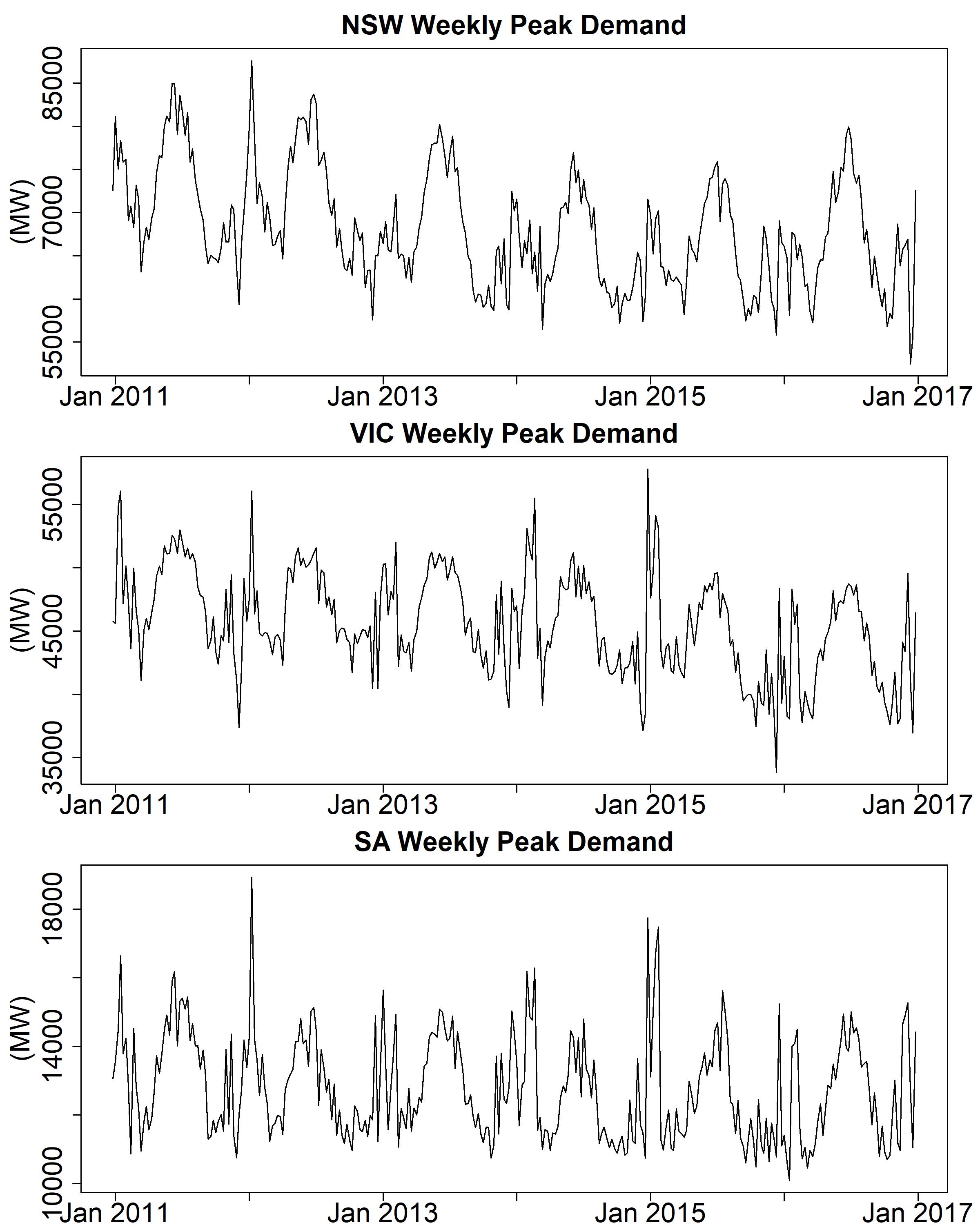}
		\caption{Time series of the aggregated WPD for NSW, VIC, and SA over all of the training data. For brevity and clarity other graphs in this report will only show the last three years of training data.}
		\label{PowerDemand}
	\end{center}
\end{figure}

Figure \ref{PowerDemand} displays the time series of WPD from $2014$ to $2016$ (inclusive). Previous years show similar seasonal trends. Visual inspection of these graphs reveals that the seasonal trends may vary between the states.

\begin{remark}
	All data analysis and graphing are conducted in R using the packages ``astsa''\footnote{\url{cran.r-project.org/web/packages/astsa/index.html}}, ``forecast'' \footnote{\url{cran.r-project.org/web/packages/forecast/index.html}}, and ``tseries''\footnote{\url{cran.r-project.org/web/packages/tseries/index.html}}, .
\end{remark}

%------------------------------------------------------
\section{Crude \texttt{SARIMA} Model: WPD Time Series}
\label{SecSARIMA}
%------------------------------------------------------

We start this section by introducing a formal definition of a \texttt{SARIMA} model. 

\begin{definition}[\cite{SS2010}]
	A time-series $\{ x_t;\, t=0,1,\ldots \}$ is \\ $\mathtt{SARIMA}(p,d,q)\times(P,D,Q)_S$,
	if
	\begin{align*}
	\Phi_P\left(B^S\right)\phi(B)\nabla^D_S\nabla^dx_t & = \delta + \Theta_Q\left(B^S\right)\theta(B)w_t,
	\end{align*}
	where $\{ w_t;\, t=0,1,\ldots \}$ is a Gaussian white noise series, $B$ is the backshift operator (i.e., $B^k x_t = x_{t-k}$), and
	\normalsize{
		\begin{align*}
		\phi(B) & = 1-\phi_1 B - \phi_2B^2 -\dots - \phi_pB^p,  \\
		\Phi_P\left(B^S\right) & = 1-\Phi_1 B^S - \Phi_2B^{2S} -\dots - \Phi_PB^{PS}, \\
		\theta(B) & = 1+\theta_1B+\theta_2B^2+\dots+\theta_qB^q,  \\
		\Theta_Q\left(B^S\right) & = 1+\Theta_1B^S+\Theta_2B^{2S}+\dots+\Theta_QB^{QS}, \\
		\nabla^d & = (1-B)^d, \\
		\nabla^D_S & = (1-B^{S})^D.
		\end{align*}}
\end{definition}
The autoregressive order $p$, moving average order $q$, seasonal autoregressive order $P$, seasonal moving average order $Q$, differencing orders $d$ and $D$, seasonal lag $S$, autoregressive coefficients $\phi_i$, moving average coefficients $\theta_i$, seasonal autoregressive coefficients $\Phi_i$, seasonal moving average coefficients $\Theta_i$, and the intercept $\delta$ are unknown parameters and should be estimated. 

Box and Jenkins \cite{BOX2015} showed that if a time series was non-stationary due to a trend in the mean, it could be detrended and converted to a stationary time series by differencing at appropriate lag(s). Perhaps, this is the main contribution of the \texttt{SARIMA} model in theory and practice. 

Intuitively, ``stationarity'' means that the statistical properties of a time series do not vary over time. More precisely, a time series is stationary, if the mean function is constant (with respect to time), and the autocovariance function for two observations of the series depends only on the time difference, the so-called \emph{lag},  between two observation points, not the actual times. A common statistical test to investigate such property for a given time series is the ``Kwiatkowski-Phillips-Schmidt-Shin'' (KPSS) test with the following hypotheses \cite{KPSS1992}:
\[\begin{cases}
H_0 : \mbox{ The time series is stationary.} \\
H_A : \mbox{ The time series is not stationary.}
\end{cases}\]

After implementing the KPSS test on the aggregated WPD data for the three states NSW, VIC and SA, it is reveled that the p-values of all of them are less than $0.01$, implying that the null hypothesis is rejected at a significance level of $1\%$. Thus, all three WPD time series are not stationary. However, we estimate an appropriate differencing orders $d$ and $D$ and the seasonality lag $S$ for each time series to convert them to a stationary time series. The outcomes of the KPSS test on before and after differenced time series are provided in Table \ref{KPSSTab}.
\begin{table}[h!]
	\centering
	\caption{The KPSS test p-values for time series before and after differencing along with the estimated values of $d$, $D$, and $S$.}
	\label{KPSSTab}  
	\begin{tabular}{llcc}
		\hline\noalign{\smallskip}
		State & Before & After \\
		\noalign{\smallskip}\hline\noalign{\smallskip}
		NSW & $< 0.01$ & $0.10$ ($d=1$, $D=1$ and $S=52$) \\
		VIC & $< 0.01$ & $0.10$ ($d=0$, $D=1$ and $S=52$) \\
		SA & $< 0.01$ & $0.10$ ($d=0$, $D=1$ and $S=52$) \\
		\noalign{\smallskip}\hline
	\end{tabular}
\end{table}

To assist in choosing the order parameters for the model, including $p$, $q$, $P$, and $Q$, the autocorrelation and partial autocorrelation plots are applied. They would come up with a few options for the orders. Ultimately, the best model (i.e., set of orders) is selected by finding the set achieving the minimum AICc (corrected Akaike information criterion) \cite{AICc1989}. AICc-based model choice enables us to balance the model complexity with the model ability to extract information from the training data \cite{BORO2017}. Furthermore, we restrict the maximum sum of orders (i.e., $p + q + P + Q$) to five to balance the model accuracy with complexity. As a final check, all coefficient p-values were assessed to be significant. The final fitted models and the estimated parameters along with their corresponding p-values are presented in Tables \ref{SARIMA} and \ref{SARIMAPara}, respectively.
\begin{table}[h!]
	\centering	
	\caption{Estimated the \texttt{SARIMA} model orders}
	\label{SARIMA}       % Give a unique label
	\begin{tabular}{lllllll}
		\hline\noalign{\smallskip}
		State & p & d & q & P& D& Q   \\
		\noalign{\smallskip}\hline\noalign{\smallskip}
		NSW & 1 & 1 & 3 & 1 & 1 & 1 \\
		VIC & 2 & 0 & 0 & 1 & 1 & 0 \\
		SA  & 1 & 0 & 0 & 0 & 1 & 0 \\
		\noalign{\smallskip}\hline
	\end{tabular}
\end{table}

\begin{table}[h!]
	\centering	
	\caption{The estimates of \texttt{SARIMA} parameters for the crude model with their p-values in brackets underneath.}
	\label{SARIMAPara}
	\begin{tabular}{llllllll}
		\hline\noalign{\smallskip}
		State & AR1 & AR2 & MA1 & MA2 & MA3 & SAR1 & SMA1   \\
		\noalign{\smallskip}\hline\noalign{\smallskip}
		NSW &  \vtop{\hbox{\strut-0.81} \hbox{\strut(0.00)}} &	& \vtop{\hbox{\strut0.24} \hbox{\strut(0.02)}} 	&\vtop{\hbox{\strut-0.79} \hbox{\strut(0.00)}} &\vtop{\hbox{\strut-0.26} \hbox{\strut(0.00)}}  &\vtop{\hbox{\strut-0.43} \hbox{\strut(0.00)}}	& \vtop{\hbox{\strut-0.32} \hbox{\strut(0.00)}}	\\
		VIC & \vtop{\hbox{\strut0.25} \hbox{\strut(0.00)}} &	\vtop{\hbox{\strut0.15} \hbox{\strut(0.00)}} & &	 &	 &		 \vtop{\hbox{\strut-0.54} \hbox{\strut(0.00)}} &	\\
		SA & \vtop{\hbox{\strut0.24} \hbox{\strut(0.00)}} &	 &	 &	 &		 & \\
		\noalign{\smallskip}\hline
	\end{tabular}
\end{table}

%-------------------------------------------------------------------------
\section{Hybrid \texttt{SARIMA}-Regression Model: Environmental Influence}
\label{SecWV}
%-------------------------------------------------------------------------

\begin{figure}[h!]
	\begin{center}
		\includegraphics[scale=0.50]{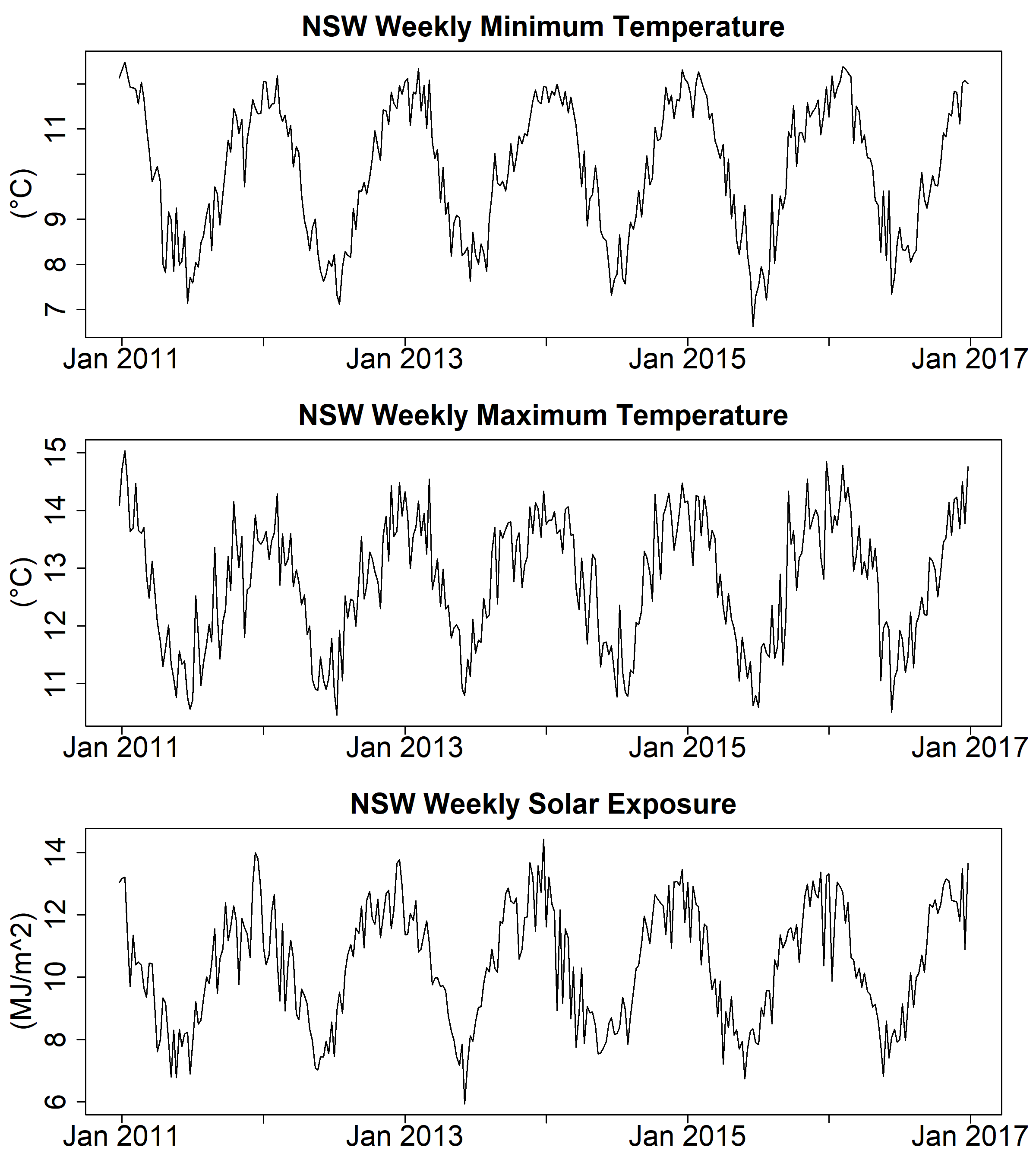}
		\caption{Maximum temperature, minimum temperature and solar exposure time series for NSW from $2014$ to $2016$ (inclusive)}
		\label{WeatherTSPlot}
	\end{center}
\end{figure}

In order to construct an appropriate hybrid \texttt{SARIMA}-regression model, we first need to realize the relationship between the primary time series WPD and the three environmental time series, including maximum temperature ($\mathtt{Max}_t$), minimum temperature ($\mathtt{Min}_t$), and solar exposure ($\mathtt{Sol}_t$). Figure \ref{PowerDemand} demonstrates that all three WPD time series possess a strong seasonal component, appearing to vary with the location. Analogously, Figure \ref{WeatherTSPlot} displays a similar temporal and spatial variation for the secondary environmental time series (to save space, only the NSW environmental time series are displayed). This observation implies that there could potentially be a significant relationship between the primary and secondary time series. 

Since the inference theory for the hybrid \texttt{SARIMA}-regression models with stationary regressor variables is completely different form that with non-stationarity variables, we need to test the stationarity of the environmental time series data at the outset. Therefore, the KPSS test is implemented on them and the corresponding p-values are reported in Table \ref{WeatherPVal}. This table indicates that all three environmental time series over the three states are stationary at a significance level of $1\%$. Indeed, this outcome is visually supported by Figure \ref{WeatherTSPlot}.
\begin{table}[h!]
	\centering	
	\caption{The KPSS test p-values for the environmental time series data.}
	\label{WeatherPVal}
	\begin{tabular}{llll}
		\hline\noalign{\smallskip}
		Environmental Variable & NSW & VIC  & SA    \\
		\noalign{\smallskip}\hline\noalign{\smallskip}
		$\mathtt{Max}_t$ &  0.12 &	0.06 &	0.05 	\\
		$\mathtt{Min}_t$ & 0.05 &	0.05 & 0.04	\\
		$\mathtt{Sol}_t$ & 0.08 &	0.04 & 0.04	  \\
		\noalign{\smallskip}\hline
	\end{tabular}
\end{table}

\begin{figure}[h!]
	\begin{center}
		\includegraphics[scale=0.55]{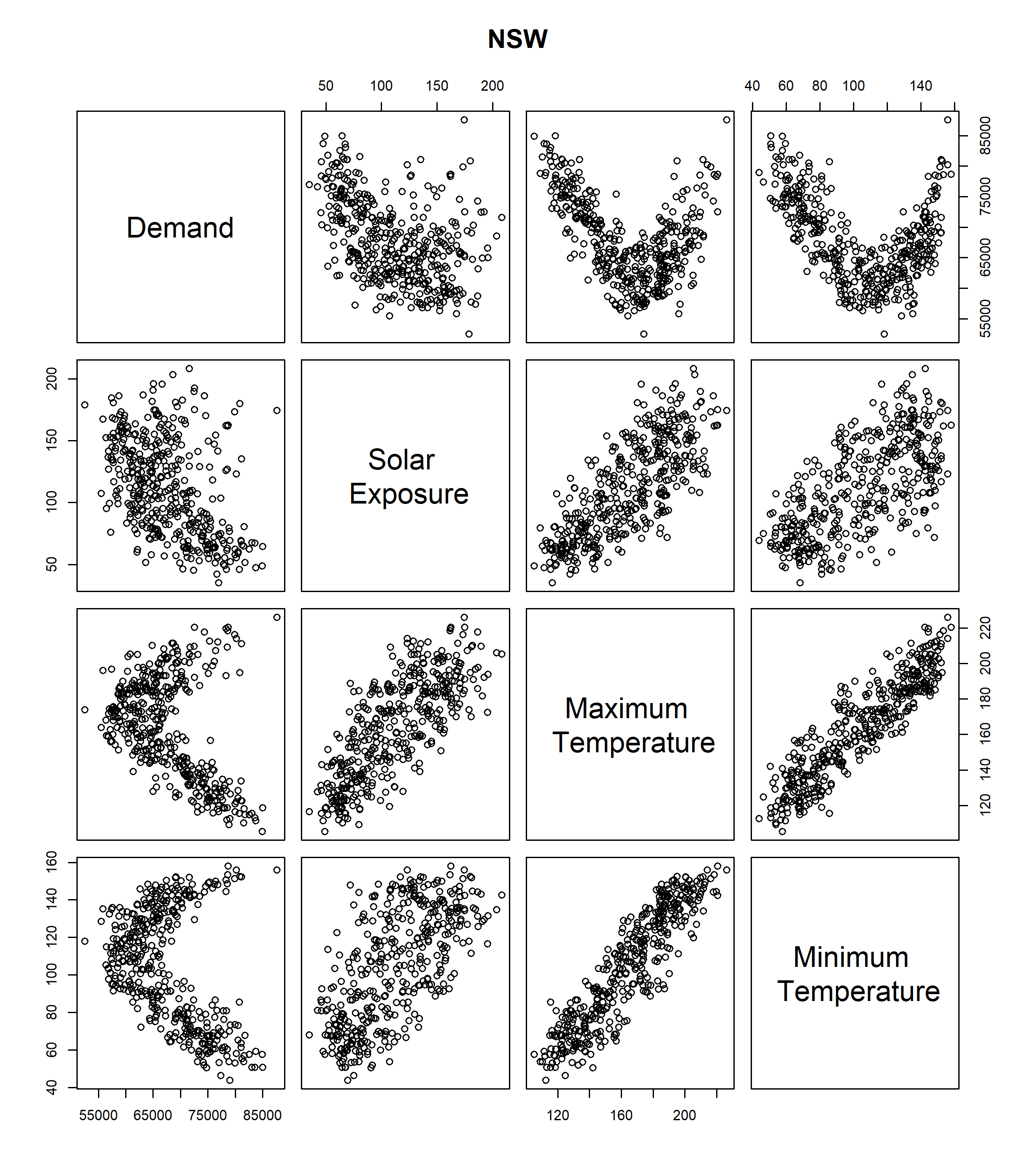}
		\caption{Scatter plot for the NSW data showing the presence of quadratic trends between WPD and the environmental variables}
		\label{Weather}
	\end{center}
\end{figure}

To investigate possible relationships between these exogenous environmental time series and the primary WPD time series, scatter plots are utilized. Figure \ref{Weather} displays the scatter plots for NSW. This figure suggests that while the maximum and minimum temperatures have a strong quadratic relationship with the WPD data, such relationship may not be as strong for the solar exposure. 

These observations would suggest $27$ combinations of the environmental variables (none, linear, and quadratic for each variable) for the ``regression'' component of the hybrid model. Once again, AICc is used to find the best combination, taking into account the secondary time series data.

The significance of each coefficient of the AICc chosen model was assessed and the final selected combinations are presented in Table \ref{WV}. This table shows that, while NSW and VIC require the full group of regression variables, surprisingly, SA does not seem to obtain sufficient benefit from the solar exposure time series. The estimates of model parameters with their corresponding p-values are presented in Tables \ref{PR1} and \ref{PR2}.
\begin{table}[h!]
	\centering	
	\caption{Selected combination of environmental variables based on the minimum value of AICc for each state}
	\label{WV}
	\begin{tabular}{ll}
		\hline\noalign{\smallskip}
		State & Schematic structure of the regression component \\
		\noalign{\smallskip}\hline\noalign{\smallskip}
		NSW &  $\mathtt{Max}_t + \mathtt{Max}_t^2 +\mathtt{Min}_t + \mathtt{Min}_t^2 +\mathtt{Sol}_t  + \mathtt{Sol}_t^2$  \\
		VIC & $\mathtt{Max}_t + \mathtt{Max}_t^2 +\mathtt{Min}_t + \mathtt{Min}_t^2+ \mathtt{Sol}_t  + \mathtt{Sol}_t^2$  \\
		SA & $\mathtt{Max}_t + \mathtt{Max}_t^2 +\mathtt{Min}_t + \mathtt{Min}_t^2 $ \\
		\noalign{\smallskip}\hline
	\end{tabular}
\end{table}

\begin{table}[h!]
	\centering	
	\caption{The estimates of \texttt{SARIMA} parameters for the hybrid model with their p-values in brackets underneath.}
	\label{PR1}
	\begin{tabular}{lllllll}
		\hline\noalign{\smallskip}
		State & AR1 & AR2 & MA1 & MA2 & SAR1 & SMA1   \\
		\noalign{\smallskip}\hline\noalign{\smallskip}
		NSW &  \vtop{\hbox{\strut-0.90} \hbox{\strut(0.00)}} &	& \vtop{\hbox{\strut0.16} \hbox{\strut(0.07)}} 	&\vtop{\hbox{\strut-0.74} \hbox{\strut(0.00)}} & &\vtop{\hbox{\strut-0.54} \hbox{\strut(0.00)}} 	 	\\
		VIC & \vtop{\hbox{\strut0.48} \hbox{\strut(0.00)}} &	\vtop{\hbox{\strut0.30} \hbox{\strut(0.00)}} &	 &	 &		 \vtop{\hbox{\strut-0.46} \hbox{\strut(0.00)}} &	\\
		SA & \vtop{\hbox{\strut0.39} \hbox{\strut(0.00)}} &\vtop{\hbox{\strut0.13} \hbox{\strut(0.03)}} 	 &	 &	 &		 & \\
		\noalign{\smallskip}\hline
	\end{tabular}
\end{table}

\begin{table}[h!]
	\centering	
	\caption{The estimates of regression parameters for the hybrid model with their p-values in brackets underneath (coefficients are rounded to one decimal places for brevity).}
	\label{PR2}
	\begin{tabular}{lllllll}
		\hline\noalign{\smallskip}
		State &$\mathtt{Max}_t$& $\mathtt{Max}_t^2$ & $\mathtt{Min}_t$ & $\mathtt{Min}_t^2$ & $\mathtt{Sol}_t$& $\mathtt{Sol}_t^2$  \\
		\noalign{\smallskip}\hline\noalign{\smallskip}
		NSW &\vtop{\hbox{\strut-770.8} \hbox{\strut(0.00)}}  &\vtop{\hbox{\strut2.4} \hbox{\strut(0.00)}}	 &\vtop{\hbox{\strut-497.3} \hbox{\strut(0.00)}} &\vtop{\hbox{\strut2.6} \hbox{\strut(0.00)}}  &\vtop{\hbox{\strut-61.6} \hbox{\strut(0.03)}}	\ &\vtop{\hbox{\strut0.27} \hbox{\strut(0.01)}}	 \\
		VIC &\vtop{\hbox{\strut-328.2} \hbox{\strut(0.00)}}  &\vtop{\hbox{\strut1.2} \hbox{\strut(0.00)}}	 &\vtop{\hbox{\strut-167.7} \hbox{\strut(0.00)}}&	\vtop{\hbox{\strut1.2} \hbox{\strut(0.00)}} &\vtop{\hbox{\strut-83.6} \hbox{\strut(0.00)}} &\vtop{\hbox{\strut0.3} \hbox{\strut(0.00)}}\\
		SA &\vtop{\hbox{\strut-156.3} \hbox{\strut(0.00)}}  &\vtop{\hbox{\strut0.5} \hbox{\strut(0.00)}}	&\vtop{\hbox{\strut-82.7} \hbox{\strut(0.00)}} &\vtop{\hbox{\strut0.5} \hbox{\strut(0.00)}} &	 &	 \\
		\noalign{\smallskip}\hline
	\end{tabular}
\end{table}

\begin{figure}[h!]
	\begin{center}
		\includegraphics[scale=0.50]{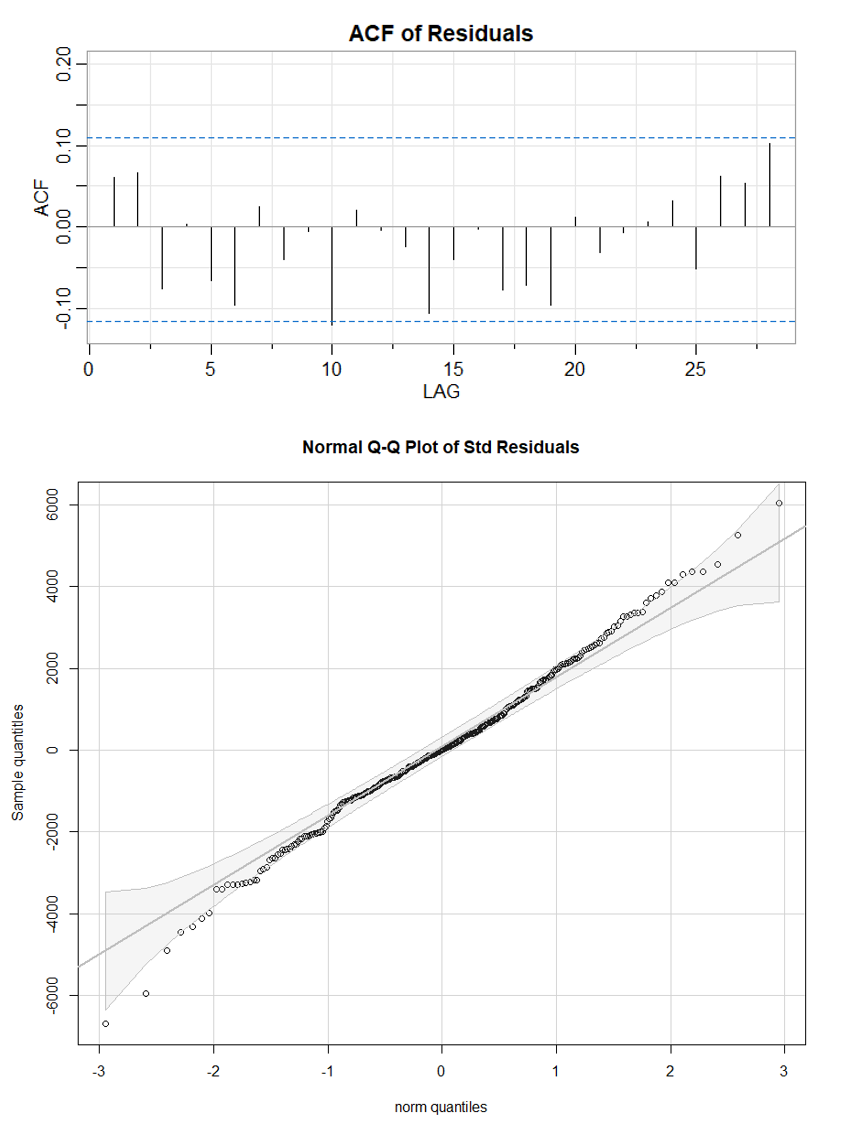}
		\caption{ACF and QQ plots for the residuals from the fitted hybrid \texttt{SARIMA}-regression model for NSW}
		\label{analysis}
	\end{center}
\end{figure}

\paragraph{Model Validation. } The estimated models are checked for statistical validity by analyzing the residuals. Figure \ref{analysis} shows the autocorrelation function (ACF) as well as QQ-plot for the residuals from the fitted hybrid \texttt{SARIMA}-regression model to the NSW WPD data. Clearly, the residuals have no autocorrelation at any lag, and the vast majority of the QQ-plot lies well within the 95\% significance area (i.e., shaded gray). Similar results are observed for the other two states.  

%--------------------------------------
\section{Medium-term Load Forecasting}
\label{SecFore}
%--------------------------------------

\begin{figure}[h!]
	\begin{center}
		\includegraphics[scale=0.35]{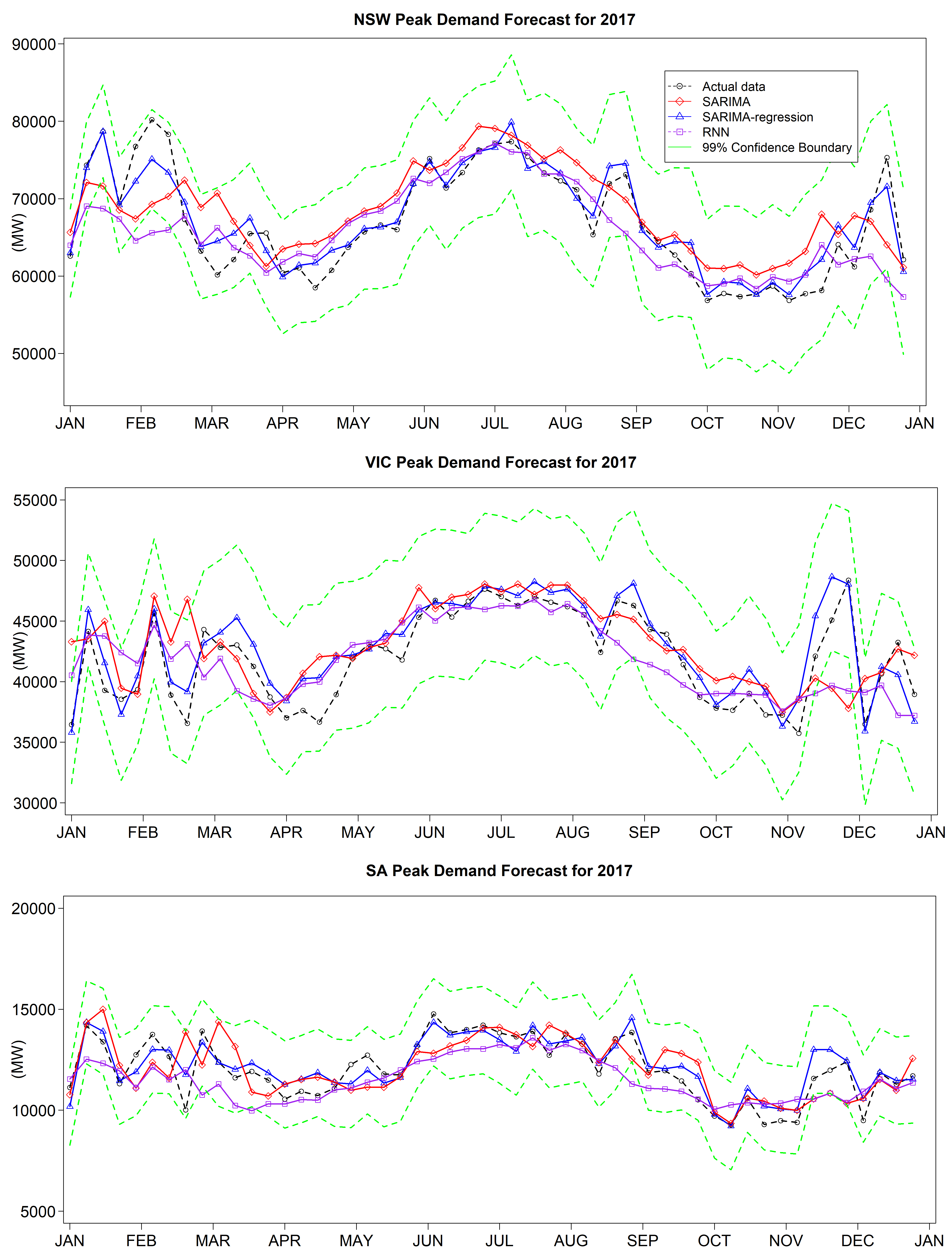}
		\caption{Comparison of the forecasts for the \texttt{SARIMA} and \texttt{SARIMA}-regression models to the actual WPD data for $2017$}
		\label{fig:c1}
	\end{center}
\end{figure}

The two crude \texttt{SARIMA} and hybrid \texttt{SARIMA}-regression models constructed in Sections \ref{SecSARIMA} and \ref{SecWV} are used to predict the WPD for all three states over $52$ weeks in $2017$. The results are displayed in Figure \ref{fig:c1}. In this figure, the black, red, blue and green plots are actual demands, forecasts generated by the \texttt{SARIMA} model, forecasts generated by the \texttt{SARIMA}-regression model, and the $99\%$ confidence boundary for WPD, respectively.  

It is readily seen that the \texttt{SARIMA}-regression model performs significantly better than the \texttt{SARIMA} model. A more solid comparison can be carried out by finding the following two popular measures to assess the effectiveness of the forecasts.

\begin{definition}[\cite{MAE}]\label{Def_MAPE}
	The \emph{mean absolute error} (MAE) is defined as:
	\begin{align*}
	\text{MAE} & = \frac{\sum_{t=1}^{h}\mid f_t - x_t\mid}{h},
	\end{align*}
	where  $f_t$, $x_t$ and $h$ are the forecast values, actual values, and prediction horizon, respectively. Analogously, the \emph{mean absolute percentage error} (MAPE) is given by  
	\begin{align*}
	\text{MAPE} & = \frac{\sum_{t=1}^{h}\left\lvert\frac{  f_t-x_t }{x_t}\right\rvert}{h}\times 100\%.
	\end{align*}	
\end{definition} 

Tables \ref{MAE} and \ref{mape}   display MAE and MAPE for the two estimated models and show the percentage improvement by employing the exogenous environmental time series into the model. The MAE and MAPE suggest an average $46.6\%$ and $46.3\%$ improvement in the accuracy of forecasts when the environmental factors are included in the model, respectively. These observations highly support the \emph{importance of environmental factors} in forecasting Australian peak power demand. 
\begin{table}[h!]
	\centering	
	\caption{Comparison of MAE for the \texttt{SARIMA} and \texttt{SARIMA}-regression models}
	\label{MAE}       % Give a unique label
	\begin{tabular}{llll}
		\hline\noalign{\smallskip}
		State & \texttt{SARIMA} & \texttt{SARIMA}-regression & Improvement (\%) \\
		\noalign{\smallskip}\hline\noalign{\smallskip}
		NSW	& 3962 & 1643 &	58.5\\
		VIC & 2225 & 1372 &	38.3\\
		SA  & 885 &	504 &	43.0\\
		\noalign{\smallskip}\hline
	\end{tabular}
\end{table}

\begin{table}[h!]
	\centering	
	\caption{Comparison of MAPE for the \texttt{SARIMA} and \texttt{SARIMA}-regression models}
	\label{mape}       % Give a unique label
	\begin{tabular}{llll}
		\hline\noalign{\smallskip}
		State & \texttt{SARIMA} & \texttt{SARIMA}-regression & Improvement (\%) \\
		\noalign{\smallskip}\hline\noalign{\smallskip}
		NSW	& 5.98 & 2.48 & 58.6 \\
		VIC & 5.49 & 3.37 & 38.6 \\
		SA  & 7.48 & 4.38 & 41.7 \\
		\noalign{\smallskip}\hline
	\end{tabular}
\end{table}

\paragraph{Machine learning approach. } In order to compare the performance of our proposed models with other methods, we apply the state-of-the-art machine learning approach to forecast WPD. More precisely, we use Recurrent Neural Networks (RNN) based GFM proposed by \cite{Hewamalage2019-il}. Table \ref{GFM_errors} summarises the optimal hyper-parameter values used in our experiments. According to \cite{Hewamalage2019-il}, these optimal hyper-parameters are determined by a sequential model-based algorithm configuration (SMAC), a variant of Bayesian Optimisation proposed by \cite{Hutter2011-il}. Furthermore, this framework uses COntinuous COin Betting (COCOB) optimisation algorithm proposed by \cite{Orabona2017-il} that does not require tuning of the network learning rate.

\begin{table}[h!]
	\centering	
	\caption{The hyper-parameter values used to train the GFM based RNN.}
	\label{hyper}       % Give a unique label
	\begin{tabular}{ll}
		\hline\noalign{\smallskip}
		Model Parameter & Optimimal Parameter value \\
		\noalign{\smallskip}\hline\noalign{\smallskip}
		RNN cell dimension	&24 \\
		Mini-batch size & 1\\
		Epoch size &3\\
		Maximum epochs &38\\ 
		Hidden layers &2\\
		Gaussian noise injection & $2 \times 10^{-4}$\\
		Random-normal initialiser & $2 \times 10^{-4}$\\
		L2-regularisation weight & $6 \times 10^{-4}$\\
		\noalign{\smallskip}\hline
	\end{tabular}
\end{table}

The MAE and MAPE of forecasts generated by this method are reported in Table \ref{GFM_errors}. We observe that the hybrid \text{SARIMA}-regression model thoroughly outperforms the GFM benchmark.
\begin{table}[h!]
	\centering	
	\caption{The MAE and MAPE for the RNN based GFM}
	\label{GFM_errors}       % Give a unique label
	\begin{tabular}{lll}
		\hline\noalign{\smallskip}
		State & MAE & MAPE \\
		\noalign{\smallskip}\hline\noalign{\smallskip}
		NSW	& 3497 & 5.07 \\
		VIC & 2194 & 5.30\\
		SA  & 950 &	7.81\\
		\noalign{\smallskip}\hline
	\end{tabular}
\end{table}

\begin{remark}
	Note that while the \texttt{SARIMA}-regression model outperforms the RNN method, the former is simpler to compute and the coefficients are more easily interpreted. In practical applications, easily compared model coefficients and specifications are highly desirable. It is also noteworthy to mention that an unrolled RNN in time resembles to a  nonlinear approximation of ARMA models, which can be expressed as a NARMA(p,q) model. Here, $p$ denotes the order of lags in the autoregressive model and $q$ denotes the order of error terms in the moving average model. For more detailed comparisons between RNN and ARIMA models, we refer to \cite{Bandara2019-iv}
\end{remark}

%----------------------------------
\section{Discussion and Conclusion}
\label{SecCon}
%----------------------------------

To the best of our knowledge, this work is the first attempt to investigate the crucial role of environmental factors in the dynamics of the Australian electricity power demand. More precisely, we developed a \texttt{SARIMA}-regression model for the weekly power demand in three major states of Australia, and empirically demonstrated the significant influence of environmental factors on predictions over a medium-term load forecasting time-scale (i.e., $52$ weeks). The results revealed that while the \texttt{SARIMA}-regression model generated, on average, an MAPE of $3.41\%$ over all states, the environmental factors could improve the accuracy of forecasts by a factor of $46.3\%$. Such an excellent MAPE is comparable with the other methods listed in Section \ref{SecLtR}. However, a direct comparison might not be fair (in favor of our model) due to the lack of other MTLF studies in the literature of Australian weekly peak power demand. This highlights the potential explanatory influence and impact environmental variables may have on power demand. Furthermore, we compared our model with the state-of-the-art machine learning methods in forecasting and demonstrate the superiority of the former model.

The weather regression variables used within this work are historical data and provided without forecasting. This was done to maximise the predictive value of the regressors to highlight their importance to predicting power demand. To move the model towards practical use future work could forecast the weather variables and use the predictions for the SARIMA regression. While this is expected to reduce the accuracy of the prediction, observation shows the weather variables are strongly seasonal and stationary and so should maintain the majority of their predictive power. 

An alternative to using environmental data derived from a single weather station would be to take the data from several sites across each state with different characteristics, and then use a weighted average by population. This method may help decision makers to identify a trend in demand that could improve the modeling of WPD. A practical drawback of this method is that many weather stations do not report complete data. Hence, the regression system will have to adjust the missing values which may bring more errors into the model.

Our model provides a scaffold for future work in improving the accuracy and utility of forecasts. Incorporating additional environmental explanatory factors such as humidity and wind direction/strength could further improve the model and, consequently, the accuracy of forecasts.


\end{document}